\documentclass[prb,aps,twocolumn,groupedaddress]{revtex4}
\usepackage{graphicx}

\begin{document}

\title{Electrical transport properties of manganite powders under pressure}

\author{M. G. Rodr\'{\i}guez}
\affiliation{Laboratorio de Bajas Temperaturas, Departamento de
F\'{\i}sica, FCEyN, UBA, and IFIBA (CONICET), Ciudad Universitaria,
(C1428EHA) Buenos Aires, Argentina}
\author{A. G. Leyva}
\affiliation{Laboratorio de Bajas Temperaturas, Departamento de
F\'{\i}sica, FCEyN, UBA, and IFIBA (CONICET), Ciudad Universitaria,
(C1428EHA) Buenos Aires, Argentina}
\author{C. Acha $^1$}
\thanks{Corresponding author: acha@df.uba.ar}
\affiliation{Laboratorio de Bajas Temperaturas, Departamento de
F\'{\i}sica, FCEyN, UBA, and IFIBA (CONICET), Ciudad Universitaria,
(C1428EHA) Buenos Aires, Argentina} x

\begin{abstract}
We have measured the electrical resistance of micrometric to
nanometric powders of the La$_{5/8-y}$Pr$_y$Ca$_{3/8}$MnO$_3$ (LPCMO
with y=0.3) manganite for hydrostatic pressures up to 4 kbar. By
applying different final thermal treatments to samples synthesized
by a microwave assisted denitration process, we obtained two
particular grain characteristic dimensions (40 nm and 1000 nm) which
allowed us to analyze the grain size sensitivity of the electrical
conduction properties of both the metal electrode interface with
manganite (Pt / LPCMO) as well as the intrinsic intergranular
interfaces formed by the LPCMO powder, conglomerate under the only
effect of external pressure. We also analyzed the effects of
pressure on the phase diagram of these powders. Our results indicate
that different magnetic phases coexist at low temperatures and that
the electrical transport properties are related to the intrinsic
interfaces, as we observe evidences of a granular behavior and an
electronic transport dominated by the Space Charge limited Current
mechanism.

\end{abstract}

\maketitle

\section{Introduction}
\label{Intro}

Up to now there have been many studies trying to shown which is the
particular influence of grain size on the properties of ceramic
manganites.~\cite{Rodo96,Dutta03} It is clear that, by reducing the
grain size of the sample, the surface to volume ratio increases,
having clear implications in their physical properties. Considering
nearest neighbors, magnetic interactions are different in bulk than
in the surface layer. Also impurities and oxygen content as well as
the crystal structure can also be different, modifying particularly
transport properties for their large sensitivity to the intergrain
coupling. Finally, the phase transitions may be modified, affecting
particularly most of the physical properties of these phase
separation systems. Regarding transport properties, it is well known
fact that manganites in the paramagnetic state (PM) have a
semiconducting-like temperature dependent resistance. On the other
hand, if we are dealing with samples in powder form, only held
together by the application of external pressure (P), we would
expect that this granularity will dominate the behavior of the
electrical conduction, by determining an extrinsic mean free path of
carriers or the tunneling mechanism between grains. In this paper we
present resistance measurements as a function of temperature of
La$_{5/8-y}$Pr$_y$Ca$_{3/8}$MnO$_3$ (LPCMO with y=0.3) powders with
two distinct grain sizes (1 $\mu$m and 40 nm) Thus, by comparison,
we want to explore the influence of nanostructuration on the
electrical transport properties, to determine the main mechanism
that regulates the conductivity between grains and to determine the
sensitivity to external pressure of the magnetic ordering of this
compound.

\section{Experimental}
\label{Expe}

In this paper we study the ceramic manganite
La$_{5/8-y}$Pr$_y$Ca$_{3/8}$MnO$_3$ (LPCMO with y=0.3), whose
morphology corresponds to a non-sintered powder. These samples were
synthesized following a microwave assisted denitration
process~\cite{Leyva04} and a final heat treatment at different
temperatures to obtain grain sizes nearly two orders of magnitude
different: Sample A, with a final sintering treatment of 4 hours at
1400 $^{\circ}$C, with a grain size diameter ($D_g$) of 1-2 microns
and Sample B, treated at 800 $^{\circ}$C for 10 minutes with $D_g$
close to 40 nm. From room temperature down to 50 K this compound
presents several magnetic transitions and a complicated phase
separation scenario. A magnetic characterization and the effects of
hydrostatic pressure on the different magnetic orderings can be
found in a previous work.~\cite{Acha06}

\begin{figure} [h]
\centerline{\includegraphics[angle=0,scale=0.25]{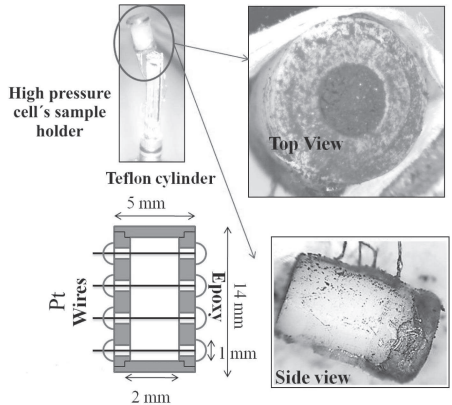}}
\vspace{-4mm} \caption{(color online) Setup to measure the
resistance of powder samples under hydrostatic pressure. Left panel:
Part of the inner sample holder of the hydrostatic cell with the
Teflon cylinder on top and its sketch. Right panel: Different views
of the Teflon cylinder filled with the LPCMO powders.} \vspace{0mm}
\label{fig:pm}
\end{figure}

In order to study the electrical transport properties of the powders
at different pressures two different techniques were used: for small
pressures (P<0.4 kbar) the powders were introduced inside a 2 mm
diameter pyrophyllite ring with Pt wires, the whole arranged in a
Bridgman-like configuration, where a small uniaxial pressure was
applied just to ensure the electrical conductivity of the powder.
For higher pressures, the inner volume of a plastic (Teflon)
cylinder was filled with the sample's powder (shown in
Fig.~\ref{fig:pm}) and pressurized inside a piston-cylinder
hydrostatic cell. The plastic cylinder has two removable lids and
four equidistant holes through its walls, ranging from side to side
along its diameter, where Pt leads were introduced to perform the
resistance measurements using a standard four terminal DC technique.
To avoid overheating and non-linear effects the exciting current was
regulated to keep the voltage drop across the sample below 10 mV. IV
characteristics were also measured using the same configuration. The
cylinder and all its holes were carefully sealed by using a low
temperature resistent epoxy in order to avoid that the pressurizing
fluid (50\%-50\% kerosene and transformer oil) mixes with the sample
preventing its electrical continuity. The proper increase of
pressure inside the cylinder with increasing the pressure of the
fluid was successfully checked by measuring the electrical
resistance of a manganin wire placed inside the cylinder with
steatite powder as the inner transmitting medium. Only a small
pressure range can be studied; up to 0.4 kbar for the Bridgman-like
setup, due to the maximum allowed uniaxial stresses. For the
hydrostatic cell, a minimum pressure (in some cases up to $\simeq$ 2
kbar) was required to have low contact resistances in order to allow
the resistance measurements, and, due to the pressure cell inner
dimensions, the maximum attained pressure was 4 kbar.

\section{Results and Discussion}
\label{R&D}

\begin{figure} [h]
\centerline{\includegraphics[angle=0,scale=0.25]{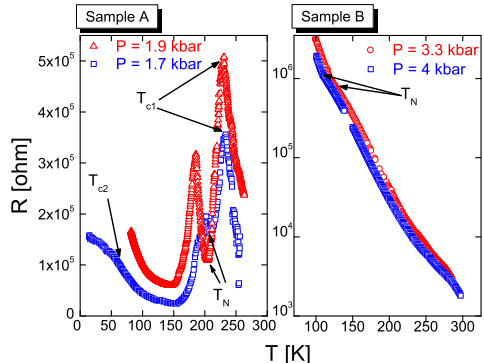}}
\vspace{-5mm} \caption{(Color online) Temperature dependence of the
resistance of Sample A and B at different pressures. The transition
temperatures are indicated for each case (T$_{c1}$, T$_N$ and
T$_{c2}$ for Sample A, and only T$_N$ for Sample B).} \vspace{0mm}
\label{fig:RvsT}
\end{figure}

The temperature dependence of resistance of both samples (A and B)
at different pressures can be observed in Fig.\ref{fig:RvsT}.
Resistance measurements shown here were taken on warming.
Measurements on cooling showed a different path due to hysteresis
effects associated with the first-order-type transition in a system
with two-phase coexistence.~\cite{Uehara99} Both samples show a high
resistance value when compared to sintered ceramic samples of
similar geometric factors.~\cite{Uehara99} Sample A shows a
semiconducting-like dependence in the paramagnetic state and very
well defined transitions to the ferromagnetic ordering and to the
charge ordered antiferromagnetic state, corresponding to the
temperatures labeled $T_{c2}, T_{c1}$ and $T_N$, respectively. A
metallic-like conduction can be observed in two intermediate
temperature regions, arbitrarily given rise to a peak in the
resistance, not necessarily associated with a magnetic order
transition but rather to the temperature dependence of a competing
distribution of metallic and insulator regions developed as a
consequence of the phase separation scenario.

As a clear difference with the resistance dependence for sintered
samples, the ground state of Sample A seems to be insulator, instead
of metallic.~\cite{Collado03} On the other hand, Sample B shows a
quite different resistance dependence, which is essentially
semiconducting-like for the whole temperature range. Nevertheless,
by analyzing the logarithmic derivative of the resistance (no shown
here), the ferromagnetic onset of $T_{c1}$ seems to be located over
300 K. At $T_N$ the resistance substantially increases, reaching
values over our experimental capacity, preventing the measurement
down to low temperatures.

\begin{figure} [h]
\centerline{\includegraphics[angle=0,scale=0.25]{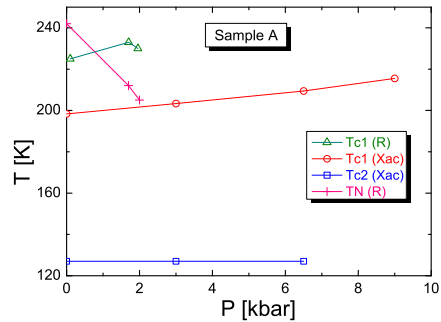}}
\vspace{-5mm} \caption{(color online) Phase diagram of Sample A. The
transition temperatures obtained previously by AC
susceptibility\cite{Acha06} are also included for comparison.}
\vspace{0mm} \label{fig:fases1400}
\end{figure}

\begin{figure} [h]
\centerline{\includegraphics[angle=0,scale=0.25]{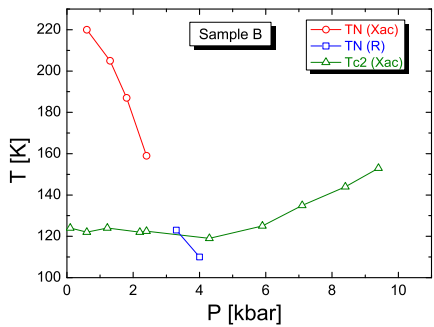}}
\vspace{-5mm} \caption{(color online) Phase diagram of Sample B. The
transition temperatures obtained previously by AC
susceptibility\cite{Acha06} are also included for comparison. There
were not evidences in the resistivity of the transitions at $T_{c1}$
and $T_{c2}$.} \vspace{0mm} \label{fig:fases800}
\end{figure}

In Fig.\ref{fig:fases1400} and Fig.\ref{fig:fases800} the obtained
phase diagram can be seen. We also included for comparison the
curves obtained in a previous paper~\cite{Acha06} where we studied
the pressure sensitivity of the magnetic ordering transitions by AC
susceptibility. The transition temperatures measured by different
techniques not necessarily coincides, but the general trend is
maintained. In this sense, the CO-AF transition is quickly reduced
by increasing the pressure while the FM transition increases softly.
Here, by measuring the pressure sensitivity of the resistance, we
confirm these previous results. Interestingly, from the magnetic
studies, the FM transition developed at $T_{c2}$ completely mask our
capacity to detect the existence of the AF transition in the case
that $T_N \leq T_{c2}$. In contrast, the resistance measurements
reveal that the CO-AF phase is still present up to 4 kbar, even for
temperatures in that range, at least for sample B. The pressure
range of our measurements should be extended to generalize these
results for both samples.

\begin{figure} [h]
\centerline{\includegraphics[angle=0,scale=0.25]{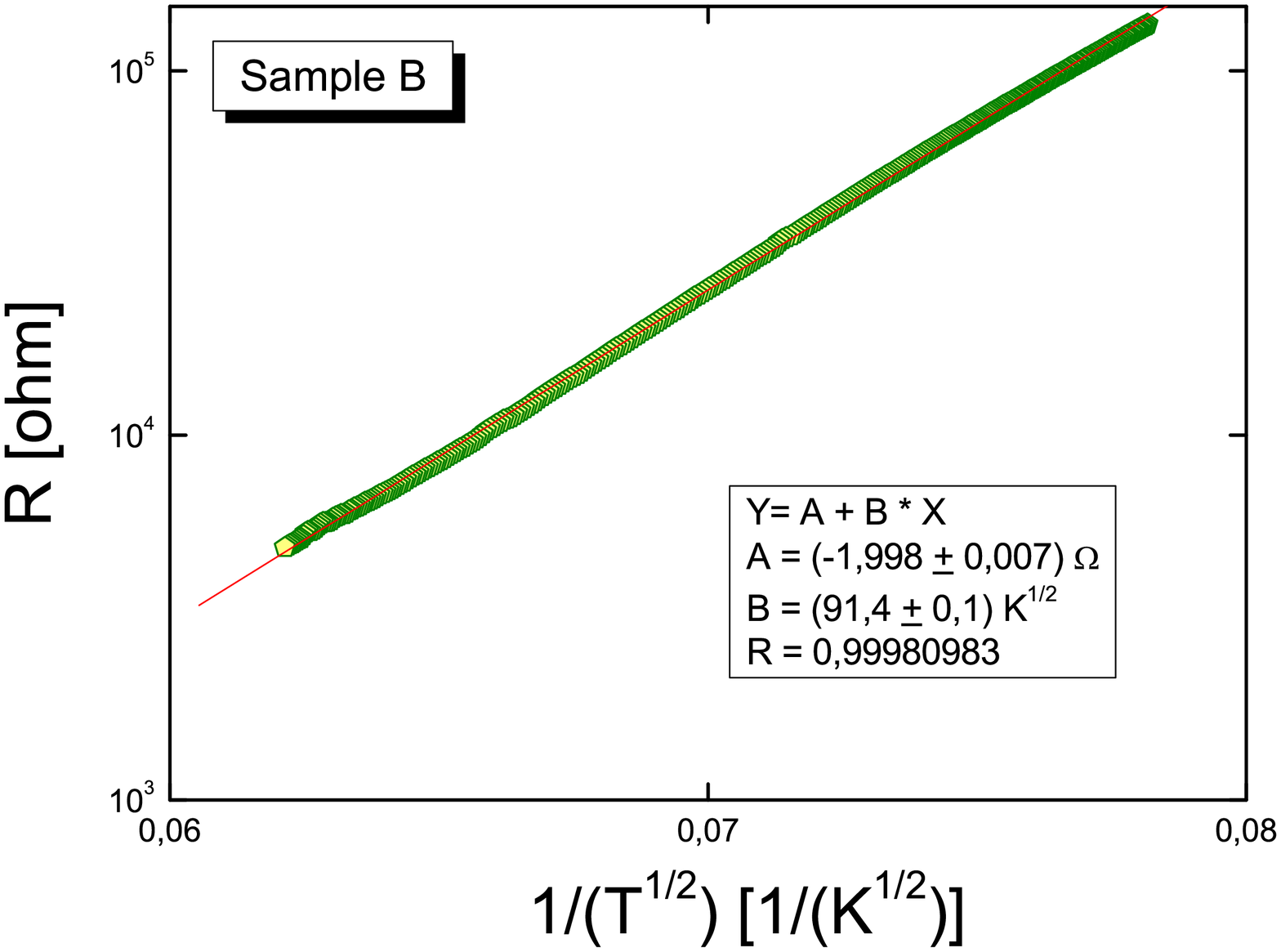}}
\vspace{-5mm} \caption{(Color online) Resistance of Sample B at 4
kbar as a function of $\frac{1}{T^{0.5}}$ in the paramagnetic
temperature range. The line is a fit using Eq.1} \vspace{0mm}
\label{fig:granular}
\end{figure}

From the temperature sensitivity of the resistance, it was not
possible to elucidate the dominant electrical transport mechanism
for Sample A, as both the fits based on a semiconductor-like
conduction or on the granular type gave reasonable results. However,
for Sample B, we obtained a better match considering the granular
conduction model~\cite{Abeles72,Abeles73} instead of the
semiconducting one, as can be observed in Fig.\ref{fig:granular}.
The expression we use to fit our data corresponds to:

\begin{equation}
\label{eq:granular} ln(R) = \frac{C_0}{T^{1/2}} = 4 \{\frac{(\chi
S^2 e^2)(1+\frac{1}{2 \chi S})}{\epsilon D(\frac{1}{2} D + S) k_b
T}\}^{1/2},
\end{equation}
\noindent where $\chi^2 = \frac{2m \Delta \phi}{\hbar^2}$, $m$ and
$e$ are the electron's mass and charge, respectively, $\Delta \phi$
the energy difference between the effective barrier height and the
electrons energy, $k_b$ the Boltzmann constant, $\epsilon$ the
dielectric constant of the insulator, $D$ its characteristic grain
size and $S$ the grain separation distance, associated with the
insulator layer between the conducting grains. A crude estimation of
$S$ can be performed, assuming that $\Delta \phi$ is in the 60-200
meV range (from the conductivity activation
energy\cite{Dagotto01,Garba04}) and $\epsilon \simeq$ 10-100.
~\cite{Cohn04} The obtained value is in the 1-10 nm range, in a very
good agreement with previous estimations.~\cite{Curiale09}

\begin{figure} [h]
\centerline{\includegraphics[angle=0,scale=0.25]{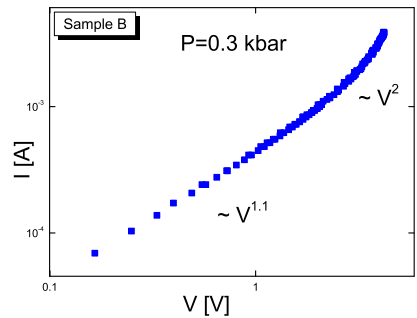}}
\vspace{-5mm} \caption{(Color online) IV characteristics at room
temperature for Sample B at 0.3 kbar. A crossover from a nearly
ohmic dependence to a $I \sim V^2$ law can be observed, typical of a
SCLC conduction mechanism at granular interfaces.} \vspace{0mm}
\label{fig:IV800}
\end{figure}

\begin{figure} [h]
\centerline{\includegraphics[angle=0,scale=0.25]{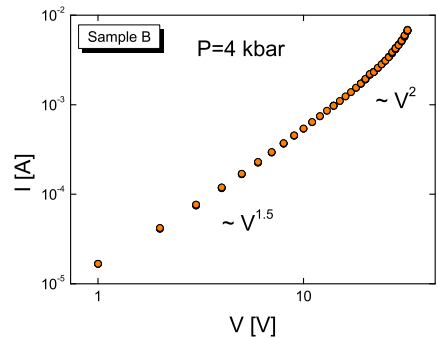}}
\vspace{-5mm} \caption{(Color online) IV characteristics at room
temperature for Sample B at 4 kbar. As in Fig.\ref{fig:IV800} the
obtained dependence is representative of a SCLC mechanism, where
electrons are injected by good conducting zones into semiconductors
in close electrical contact. Here, the crossover to a $I \sim V^2$
occurs at higher voltages indicating that pressure favors the
conduction of the semiconducting zones.} \vspace{0mm}
\label{fig:IV800conP}
\end{figure}

We also analyze the IV characteristics in order to gain insight on
the conduction mechanism between interfaces. We present in
Fig.\ref{fig:IV800} and Fig.\ref{fig:IV800conP} results for Sample
B, although a similar behavior was obtained for Sample A. In both
figures a non-linear dependence can be observed that evolves from a
nearly ohmic relation to a power law with $I \sim V^2$.

This behavior is typical of the space charge limited currents (SCLC)
mechanism, obtained for interfaces formed by good metals in contact
with semiconductors.~\cite{Dietz95} It is interesting to note that
the crossover to the quadratic dependence is shifted to higher
voltage values with increasing pressure, indicating that pressure
extend the ohmic behavior of the interface, probably by increasing
the conductivity of the semiconductor.~\cite{Joshi93} As the IV
characteristics were measured in a four terminal configuration, it
is clear that the SCLC conduction is not only related to the
interface formed by the Pt electrodes with the powder but it is
essentially associated with the intrinsic interfaces formed between
grains. As in the granular model that describes the temperature
dependent resistance, the SCLC bulk conduction is revealing the
existence of good metallic zones (grains) very well connected to a
semiconducting layer (surface of grains).

\section{Conclusions}
\label{Conc}

We have measured, as a function of temperature and pressure, the
electrical transport properties of LPCMO powders samples with grain
sizes in the micrometric to nanometric scale. Both samples show a
conduction behavior dominated by the interfaces, with
non-linearities based on the SCLC mechanism and a conduction regime
determined by their extrinsic granularity. Within the granular
model, we made a crude estimation of the insulator layer obtaining a
value in the 1-10 nm range. Considering the magnetic order
transitions, we obtain a phase diagram that confirms previous
results obtained by magnetic measurements. Moreover, new results
were obtained as we observed the coexistence of the CO-AF phase with
the low temperature FM phase confirming the complex phase diagram of
these phase separated compounds.




\end{document}